\documentclass[%
 reprint,
showpacs,
 amsmath,amssymb,
 aps,
 pra,
floatfix,
superscriptaddress
]{revtex4-1}
\usepackage[utf8]{inputenc}
\usepackage[T1]{fontenc}
\usepackage[]{graphicx}
\usepackage{color}
\usepackage[dvipsnames]{xcolor}
\usepackage[english]{babel}
\usepackage[normalem]{ulem}
\usepackage[bookmarks,bookmarksopen,bookmarksdepth=2]{hyperref}
\hypersetup{
    colorlinks=true,
    citecolor=blue,
    linkcolor=blue,
    filecolor=magenta,
    urlcolor=cyan,
}

\usepackage{microtype}
\usepackage{braket}
\usepackage{esvect}
\usepackage[export]{adjustbox}
\usepackage{dsfont}
\usepackage[T1]{fontenc}
\usepackage[utf8]{inputenc}
\usepackage{url}
\usepackage{orcidlink}

\usepackage{amsmath,amsfonts,amssymb,amsthm}
\usepackage{mathtools}
\usepackage{bbm}
\usepackage{mathrsfs,physics,xfrac,cancel,tensor,relsize,tikz,lipsum,quantikz}

\newcommand{\expv}[1]{\langle#1\rangle}

\begin{document}

\title{A Hybrid Measurement Scheme for Generating nonGaussian Spin States}
\author{Andrew Kolmer Forbes\orcidlink{0009-0003-8730-8007}}
\email{aforbes@unm.edu}
\author{Ivan H. Deutsch}
\affiliation{Center for Quantum Information and Control, Department of Physics and Astronomy, University of New Mexico, Albuquerque, New Mexico 87131, USA}

\begin{abstract}
We present a protocol for generating nonclassical states of atomic spin ensembles through the backaction induced by a hybrid measurement of light that is entangled with atoms, combining both homodyne and single photon detection.  In phase-I of the protocol we create a spin squeezed state by measuring the light's polarization rotation due to the Faraday effect in a balanced polarimeter, equivalent to a homodyne measurement.  In phase-II  we send a second probe beam through the sample and detect single photons scattered into the signal mode. Before doing so, we rotate the uncertainty bubble to increase the projection fluctuations of the measured spin component.  This increases the coupling strength between the atoms and photons and thus the rate of scattering of single photons into the signal mode.  In the ideal case, the result is a squeezed Dicke state, with substantial quantum advantage for sensing spin rotations.  We benchmark the protocol's utility in the presence of inevitable decoherence due to optical pumping using the Fisher information as a measure of quantum advantage.  We show that in the presence of decoherence, the quantum Fisher information associated with the nonGaussian mixed state we prepare is substantially larger than the classical Fisher information obtained from the standard measurement of spin rotations.   We deduce a measurement basis that is close to optimal for achieving the quantum Cram\'{e}r Rao bound in the presence of decoherence.
\end{abstract}

\maketitle

\section{Introduction}

Ensembles of atoms are a powerful platform for quantum sensing and metrology~\cite{pezze2018quantum} in applications including clocks, magnetometers, and inertial sensors.  The precision with which sensors perform is fundamentally limited by the quantum uncertainty principle. For uncorrelated atoms, quantum projection noise defines the ``standard quantum limit'' (SQL).  One can beat the SQL by employing correlated, entangled atoms. Spin squeezed states are the simplest example~\cite{kitagawa1993squeezed}, and substantial reduction in projection noise has been demonstrated in a number of experiments~\cite{kuzmich1997spin,leroux2010implementation, PhysRevLett.109.253605, julia2012dynamic, cox2016deterministic, engelsen2017bell, hemmer2021squeezing}, reaching levels that may soon have practical import in some applications~\cite{colombo2022entanglement, eckner2023realizing, robinson2024direct}.  

For large ensembles, spin squeezed states are Gaussian states, meaning the probability distribution for measurement outcomes of spin projection is a normal distribution. In order to approach the fundamental Heisenberg limit, one requires the creation of nonGaussian states. Large atomic ensembles are also well approximated by infinite dimensional quantum systems with potential application in continuous variable (CV) quantum information processing~\cite{kuzmich2003atomic}.  There too, the creation of nonGaussian states is an essential resource for universal quantum computing~\cite{ghose2007non, Andersen_2010}.

A mechanism to create the necessary entanglement between atoms in the ensemble, useful for metrology and CV quantum information processing, is through the atoms' mutual coupling to a mode of the optical field. This atom-light interface can be used the engineer a desired entangling unitary evolution such as one-axis or two-axis twisting~\cite{carrasco2022extreme} or to create entanglement between atoms induced by the measurement backaction after appropriate measurement of the light~\cite{kuzmich1997spin, cox2016deterministic,Mitchell2012,hemmer2021squeezing}. A Gaussian measurement of the light, such as homodyne detection, leads to Gaussian states of the atoms, e.g., spin squeezed states. The creation of nonGaussian states of the ensemble requires nonGaussian measurements of the light, such as photon-number counting. The backaction associated with the detection of a single photon entangled with a spin ensemble leads to a singly excited Dicke state, as studied in~\cite{vuletic_nature,vuletic_theory}.

In this work we consider a hybrid measurement scheme, including both Gaussian and nonGaussian measurements.  Our motivation is two fold.  We will see how the use of squeezing can enhance the measurement strength for single photon detections.  Secondly, through hybrid measurements, one can increase the quantum advantage for metrology, as quantified by the quantum Fisher information~\cite{braunstein1994statistical}.  Such hybrid measurement protocols are analogous to photon addition and subtraction, a long-studied tool in quantum optics with applications across quantum sensing and metrology. Photon added coherent states (PACS) are of particular interest in metrology, as they exhibit both classical and nonclassical characteristics depending on the amplitude of the coherent state \cite{Zavatta_2004}. With respect to sensing and metrology, photon addition/subtraction can be used to enhance signal-to-noise ratios in detection of faint thermal sources \cite{Parazzoli_2016}. It can also be used to prepare many different states nonclassical states including Schr\"odinger cat states \cite{Dakna_1997,Takahashi_2008,Ourjoumtsev_2009,Takase_2021} and Gottesman-Kitaev-Preskill (GKP) states \cite{Eaton_2019}.

\begin{figure*}[t]
    \centering
    \includegraphics[width=0.8\linewidth]{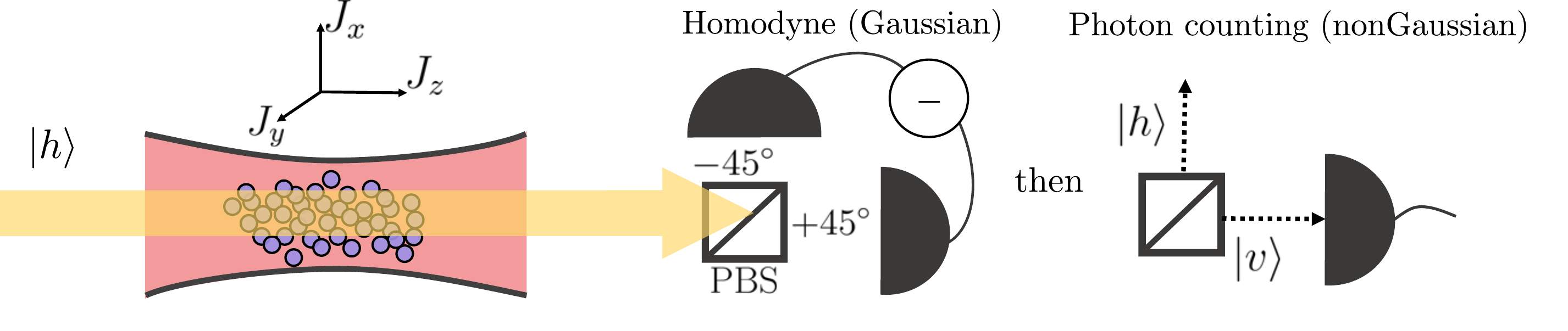}
    \caption{Schematic of the hybrid measurement protocol, with both homodyne and photon counting measurements present. On the left is an ensemble of dipole trapped atoms, with horizontally polarized probe beam passing through. After an entangling interaction under the Faraday effect, Eq. (\ref{eq:spin_unitary}), the light is measured.  In phase-I we consider a balanced polarimeter, equivalent to homodyne measurement of $\ket{v}$-polarized photons, with the probe acting as the local oscillator. In phase-II, we employ a second   $\ket{h}$-polarized pulse.  The horizontally polarized probe is filtered out, and a photon counting measurement is made on the signal in the,  $\ket{v}$-polarized mode.}
    \label{fig:protocol_sequence}
\end{figure*}

In this article we will propose a protocol which aims to generate a nonGaussian state of an atomic spin ensemble through hybrid measurements, employing the entangling atom-light interface arising from the Faraday effect~\cite{deutsch2010quantum}. Homodyne measurements correspond to polarization spectroscopy of the Faraday rotation angle~\cite{deutsch2010quantum} and discrete-variable measurement corresponds to counting photons scattered from the probe mode to the orthogonal polarization mode~\cite{vuletic_nature}.  We divide the protocol in two phases: measurement-induced spin squeezing in phase-I and the detection of a single photon in phase-II. In Sec.~\ref{sec:hybrid_measurement} we will detail this protocol, and apply previously developed theory to analytically derive equations of motion for the state. 

The ultimate utility of the hydrid measurement protocol will depend on the deleterious effect of decoherence. In the context of the atom-light interface, unavoidable decoherence occurs due to optical pumping associated with diffuse photon scattering by the atomic ensemble. We demonstrated recently that for large ensembles local optical pumping acting on spins can be represented as a bosonic decoherence channel, which makes the modeling of noise on such ensembles tractable \cite{Forbes_2024}.   We use this formalism to include decoherence in the equations of motion.

In Sec.~\ref{sec:figures_of_merit} we analyze the protocol using three figures of merit: (i) The classical Fisher information (CFI) with respect to sensing rotations by measuring the displaced spin.  This is optimal in the ideal case but not in the presence of decoherence; (ii) The quantum Fisher information (QFI) which quantifies the quantum Cram\'{e}r Rao bound in parameter estimation, achievable with the state we prepare when we measure it according to optimal POVM~\cite{braunstein1994statistical}; (iii) The total runtime of the experiment. We show that our hybrid measurement protocol can yield a substantial quantum advantage and find a POVM that comes close to achieving the QFI, and thus optimal sensing in the presence of noise.  We summarize and give an outlook in Sec. \ref{sec:conclusions}.

\section{Hybrid Measurement Protocol}
\label{sec:hybrid_measurement}

\subsection{Overview of the Protocol}
\label{sec:overview}
The basic protocol is shown in Fig.~\ref{fig:protocol_sequence}.   We consider an atom-light interface consisting of an ensemble of spin-1/2 trapped atoms coupled and mode-matched to a probe laser field.  Through the Faraday effect, the $H$-polarized probe beam passes through an atomic ensemble which scatters into $V$-polarized light. The interaction between the atomic spin and probe polarization is governed by the entangling unitary operator~\cite{deutsch2010quantum},
\begin{equation}
    \hat U=e^{-i\chi \hat J_z\otimes \hat S_3 ,}\label{eq:spin_unitary}
\end{equation}
where $\hat S_3$ is the Stokes vector component of the light corresponding to the scattering between the modes, 
\begin{equation}
    \hat{S_3} = \frac{1}{2i}(\hat{a}_H^\dag \hat{a}_V-\hat{a}_V^\dag\hat{a}_H)
\end{equation}
and $\hat J_z$ is the collective angular momentum operator along the $z$-direction,
\begin{equation}
    \hat J_z=\frac12\sum_i\hat\sigma_z^{(i)},
\end{equation}
where $\hat\sigma_z^{(i)}$ is the Pauli-$z$ operator on the $i^\mathrm{th}$ atom. The Faraday rotation angle sets the coupling strength $\chi$.

We consider large ensembles, in which the magnitude of the mean spin is much greater than the projection fluctuations.  In that case we make the Holstein-Primakoff approximation (HPA)~\cite{Primakoff_1940}, which maps spin operators to bosonic quadrature operators
\begin{align}
    \hat J_y&\to\sqrt{\frac{N_A}{2}}\hat X_A\\
    \hat J_z&\to\sqrt{\frac{N_A}{2}}\hat P_A,
\end{align}
where $N_A$ is the number of atoms in the ensemble.  The spin Dicke states are mapped to bosonic Fock states as
\begin{equation}
    \ket{J,M}\to\ket{n=J-M}.
\end{equation}
We can further apply this approximation to the components of the Stokes vector.  As the probe mode contains many photons, we can neglect its fluctuations  and set, $\hat{a}_H \rightarrow \sqrt{N_L}$, where $N_L$ is the number of photons in a temporal mode of time $T$, so that
\begin{align}
    \hat S_3\to \frac{\sqrt{N_L}}{2i}(\hat{a}_V-\hat{a}_V^\dag) =\sqrt{\frac{N_L}{2}}\hat P_L,
\end{align}
where $\hat P_L$ is the $P$-quadrature of the $V$-polarized mode.  Applying these approximations to Eq.~(\ref{eq:spin_unitary}), the unitary map takes the form of an entangling interaction between two bosonic modes,
\begin{align}
    \hat U&=e^{-i\sqrt{\frac{N_A N_L \chi^2}{4}}\hat P_A\otimes \hat P_L}\\
    &=e^{-i\sqrt{\frac{N_A\kappa T}{4}}\hat P_A\otimes \hat P_L}\label{eq:bosonic_unitary}
\end{align}
where $\kappa T= N_L \chi^2$ is the measurement strength in time $T$. Note, $N_A \kappa = OD \gamma$, where $OD$ is the optical depth on resonance, which sets to cooperativity for scattering into the probe mode, and $\gamma$ is the photon scattering rate into $4 \pi$ steradians, which determines the rate of decoherence.  Strong coupling requires $OD \gg 1$~\cite{deutsch2010quantum}. 

\begin{figure*}
    \centering
    \includegraphics[width=.85\linewidth]{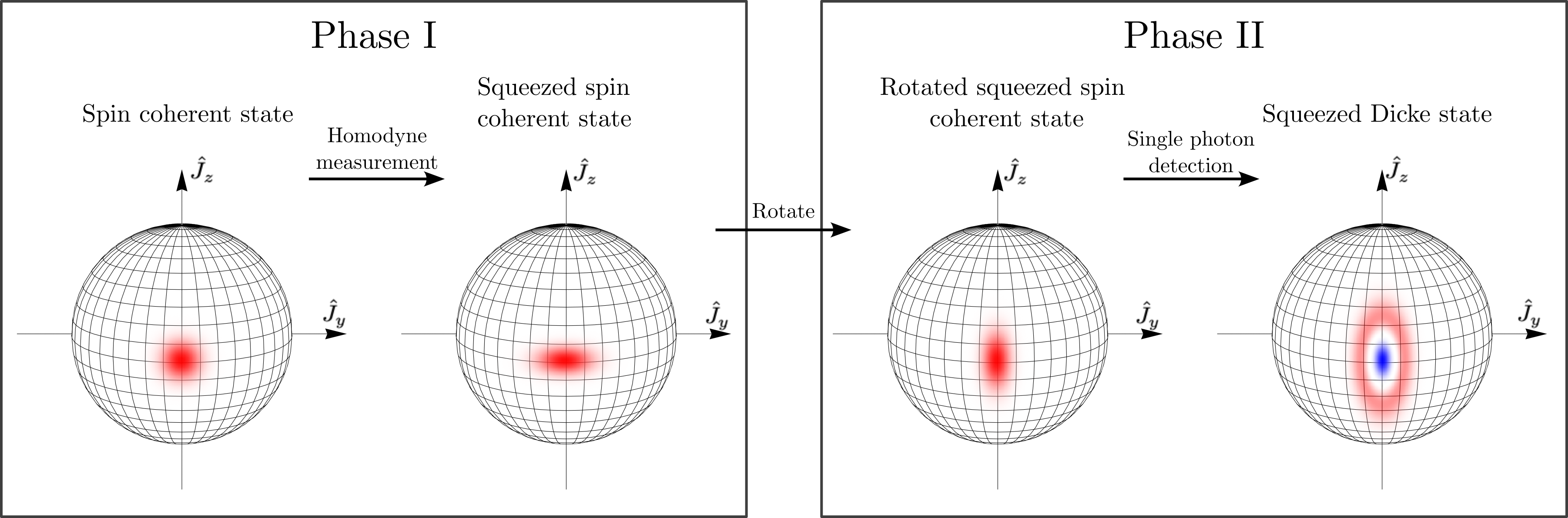}
    \caption{ Wigner function of the spin ensemble, with each box specifying a phase of the protocol. Positive regions of the Wigner function are red, while negative regions are depicted using blue. During phase-I we take a spin coherent state prepared along the $\hat J_x$ direction and measure the projection noise along the $z$-axis via the Faraday effect in a balanced polarimeter.  The the result is a spin squeezed state (when neglect the random displacement along $z$). In phase-II we rotate we rotate the spins to oriented the anti-squeezed direction along the z-axis, and then begin single photon detection. The protocol ends when a single photon is detected, generating a squeezed Dicke state.}
    \label{fig:protocol_diagram}
\end{figure*}

As discussed in the introduction, our protocol for creating nonGaussian states consists of two phases, spin squeezing followed by single photon detection. 
Prior to phase-I the ensemble is initialized in a spin coherent state along the $J_x$ direction. During the first phase we measure the scatted photons via a balanced polarimeter that corresponds to homodyne detection of the $V$-mode, with the probe field $H$-mode serving as the local oscillator~\cite{Baragiola_2014}.  The photocurrent is a measurement of the projection noise of the spin along $J_z$, which induces spin squeezing, as is shown in Fig.~\ref{fig:protocol_diagram}. 

After squeezing, the ensemble is rotated about the $J_x$ axis by $\pi/2$ to that the anti-squeezed quadrature is along $J_z$. Then in phase-II one performs single photon detection on the vertically polarized photons, similar to the heralding procedure studied in \cite{vuletic_nature}. This requires us to filter out all horizontally polarized photons from the probe beam. Once a single photon is detected, the probe is extinguished. 

In summary, the protocol consists of:
\begin{itemize}
    \item Phase-I: Prepare a spin coherent state (SCS) along the $J_x$ direction.
    \item Pass a $H$-polarized probe beam through the ensemble.
    \item Measure the Faraday rotation angle via balanced polarimeter (homodyne measurement) for time $t_1$. 
    \item Phase-II: Rotate the state about the $J_x$ axis by $\pi/2$.
    \item Perform single photon detection on the $V$-polarized light. 
    \item Extinguish the probe upon detection of a single $V$-polarized photon, which occurs at time $t_2$ after the start of phase-II.
\end{itemize}

This protocol is contrasted with protocols in which one prepares a spin coherent state and immediately begins single photon detection, skipping phase-I \cite{vuletic_nature}. The purpose of including phase-I in this protocol is two-fold. First, by squeezing the ensemble one may increase the metrological utility of the state by reducing its fluctuations along a particular axis. Second, ``anti-squeezing,'' while staying close to a minium uncertainty state, increases the effective measurement strength during single photon detection by increasing the probability of detecting a photon, as will be studied in more detail in Sec.~\ref{sec:SPD}. In the absence of a high-finesse cavity, this increase in effective measurement strength may prove useful. However, the additional steps in state preparation come at the cost of additional decoherence. One therefore needs to analyze and compare the tradeoff between more squeezing and increased decoherence, which we study in Sec.~\ref{sec:figures_of_merit}.

\subsection{Phase-I: Homodyne Measurement}
\label{sec:homodyne}

To analyze the performance of our protocol, we derive the state that is prepared through measurement backaction, including the presence of decoherence by optimal pumping. Initially, we prepare the state in a spin coherent state along the $J_x$ direction. During the first phase we apply a measurement of the $J_z$ component of the ensemble through a balanced polarimeter. This generates a spin squeezed state, with decreased fluctuations along the $J_z$ direction (equivalently the $P_A$ direction in the HPA) as seen in Fig.~\ref{fig:protocol_diagram}. In the HPA, the Kraus operator associated with the balanced polarimeter as a function of $T$ is~\cite{deutsch2010quantum}
\begin{align}
    \hat K&=\bra{X_L}e^{-i\sqrt{\frac{N_A\kappa T}{4}}\hat P_A\otimes \hat P_L}\ket{0_L}\label{eq:kraus1}\\
    &=\braket{X_L}{\alpha_L=\sqrt{\frac{N_A\kappa T}{8}}\hat P_A}\label{eq:kraus2}\\
    &=\frac{1}{\pi^{1/4}}\exp[-\frac{N_A\kappa T}{8}\left(\hat P_A-\Pi \right)^2],\label{eq:kraus3}
\end{align}
where $\ket{0_L}$ is the vacuum state in the vertical mode of light, and $\ket{X_L}$ is an eigenstate of the $\hat X_L$ quadrature of the vertically polarized mode of light, and $\Pi \equiv X_L\sqrt{4/N_A \kappa T}$ is the measurement outcome in the units of $P_A$. Homodyne measurement thus corresponds to a Gaussian Kraus operator centered at the measurement outcome and resolution $r=N_A\kappa T/2$.  The strength of the measurement is determined by resolving power of the noisy meter.  This can be re\"{e}xpressed as 
$r= \frac{(\Delta \hat J_z^2)_{PN}}{(\Delta \hat J_z^2)_{SN}}$, where $(\Delta \hat J_z^2)_{PN} = N_A/2$ is the initial spin projection noise of the state and $(\Delta \hat J_z^2)_{SN}=1/\kappa T$ is the uncertainty in the measurement outcome due photon shot noise in the probe.


Equation (\ref{eq:kraus3}), when applied to the initial state of the atoms has two effects. The first is to displace the state in the $P_A$ direction by an amount proportional to the random measured valued  $\Pi$. For a continuous weak measurement of $\hat P_A$, this leads to a stochastic evolution of $\langle\hat P_A\rangle$~\cite{Baragiola_2014}. The second effect is a deterministic squeezing of fluctuations in $P_A$. For the purposes of this analysis, the stochastic displacement does not affect the quantities of interest, and thus for the remainder of this work we will choose $\Pi=0$, and thus consider the Kraus operator in phase-I to be
\begin{equation}
    \hat K_1\equiv\exp[-\frac18 N_A \kappa T\hat P_A^2]\label{eq:K_phase1}.
\end{equation}
Given the initial SCS (vacuum in the HP approximation), the result ideal result in the HPA is a squeezed vacuum state.

In order to include decoherence, we turn to the equations of motion under optical pumping. The Lindblad master equation under this noise channel is
\begin{align}
    \frac{\mathrm d}{\mathrm dt}\hat\rho=&\gamma \sum_{i=1}^N \left(\hat\sigma_+^{(i)}\hat\rho\hat\sigma_-^{(i)}-\frac12\hat\sigma_-^{(i)}\hat\sigma_+^{(i)}\hat\rho-\frac12\hat\rho\hat\sigma_-^{(i)}\hat\sigma_+^{(i)}\right.\nonumber \\
    &\left. +\hat\sigma_-^{(i)}\hat\rho\hat\sigma_+^{(i)}-\frac12\hat\sigma_+^{(i)}\hat\sigma_-^{(i)}\hat\rho-\frac12\hat\rho\hat\sigma_+^{(i)}\hat\sigma_-^{(i)}\right)\label{eq:optical_pumping}
\end{align}
where $\gamma$ is the rate of photon scattering and $\hat\sigma_\pm^{(i)}$ are the spin-flip operators respectively on the $i^\mathrm{th}$ particle. For large $N$ one may use the HPA for the evolution of an arbitrary moment of a bosonic quadrature operator $\hat Q$ as~\cite{Forbes_2024},
\begin{equation}
    \frac{\mathrm d}{\mathrm dt}\expval{\hat Q^n}\approx -n\gamma\expval{\hat Q^n}+\frac{\gamma}{2}n(n-1)\expval{\hat Q^{n-2}},\label{eq:fp_X_evol}
\end{equation}
Note that Eq.~(\ref{eq:fp_X_evol}) is exact for $n=2$.

We consider the equations of motion of variance in $\hat X_A$ and $\hat P_A$ since this will be sufficient for defining the Gaussian state. Including both continuous-time application of $\hat K_1$ measurement backaction and optical pumping we obtain (see App.~\ref{app:x2_p2}),
\begin{align}
    \frac{\mathrm d}{\mathrm dt}\expval{\hat{X}_A^2}&=\frac{\kappa N}{8}e^{-4\gamma t}-2\gamma\expval{\hat{X}_A^2}+\gamma ,\label{eq:x2_w_dec}\\
    \frac{\mathrm d}{\mathrm dt}\expval{\hat{P}_A^2}&=-\frac{\kappa N}{2}\expval{\hat{P}_A^2}^2-2\gamma\expval{\hat{P}_A^2}+\gamma\label{eq:p2_w_dec}.
\end{align}
As the state is Gaussian with zero mean, these equations of motion completely specify the state after phase-I.

\subsection{Phase-II: Single Photon Detection}
\label{sec:SPD}

Following homodyne measurement and squeezing, in preparation of phase-II of our protocol, the spins are rotated about the $J_x$ axis by $\pi/2$ so that the anti-squeezed quadrature is now $P_A$.  This increase in fluctuations along  $P_A$ increases the probability of the ensemble producing signal photons. Physically, this can be understood because the strength of the Faraday interaction depends of the magnitude of $J_z$ (equivalently, $P_A$).  Increasing the projection noise in that quadrature increases the measurement strength strength.  Mathematically, this can be seen by considering the Kraus operator $\hat K_{\text{SPD}}$ associated with single photon detection (SPD). Conditioned on measuring a single photon in the $V$-mode,
\begin{align}
    \hat K_{\text{SPD}}&=\bra{1_L}e^{-i\sqrt{\frac{N_A\kappa T}{4}}\hat P_A\otimes \hat P_L}\ket{0_L} \nonumber \\
    &=\braket{1_L}{\alpha_L=\sqrt{\frac{N_A\kappa T}{8}}\hat P_A} \nonumber \\
    &=\exp\left\{-\frac{N_A \kappa T}{16}\hat P_A^2\right\}  \sqrt{\frac{N_A \kappa T}{8}}\hat P_A.\label{eq:KSPD}
\end{align}
where $\ket{0_L}$ and $\ket{1_L}$ are the vacuum state and single photon state respectively and $\ket{\alpha_L}$ is a coherent state all in the vertically polarized mode of light.

Equation (\ref{eq:KSPD}) implies that for some small time interval $T=\delta t$, 
\begin{equation}
    \hat K_\text{SPD}\approx\sqrt{\frac{\kappa N_A\delta t}{8}}\hat P_A\label{eq:KSPD_dt},
\end{equation}
and therefore the probability of single photon detection during a small window of time $\delta t$ is
\begin{equation}
    p_\text{SPD}=\frac{\kappa N_A\delta t}{8}\expval{\hat P_A^2}.\label{eq:p_spd}
\end{equation}
Since $p_\text{SPD}\propto\langle\hat P_A^2\rangle$, the the probability of detecting a photon is proportional to the fluctuations in the $\hat P_A$ quadrature. In this sense, phase-I serves to effectively increase the measurement strength during phase-II.

Equation (\ref{eq:KSPD}) can be interpreted as containing two types of evolution. If a single photon is detected during a small time interval $\delta t$ around time $t_2$, post-measurement state determined by the Kraus operator Eq.~(\ref{eq:KSPD_dt}), is equivalent to the application of $\hat P_A$ (followed by renormalization).  This is a nonGaussian operation corresponding to photon addition and subtraction.  In addition, the exponential decay factor in Eq. (\ref{eq:KSPD}) has the following interpretation. For all times before $t_2$ that we do {\em not} detect a photon, we must apply ``no-jump'' evolution to the state consistent with monitoring the system~\cite{molmer1993monte}. This no-jump is information we learn, and the Bayesian backaction effect reduces the spin projection noise along $P_A$; the longer we do not detect a photon the more likely there was no spin projection along $J_z$ and we must adjust our state assignment accordingly. 

The Gaussian exponential factor in Eq. (\ref{eq:KSPD}) is very similar to the squeezing Kraus operator in Eq.~(\ref{eq:K_phase1}), but with half the rate $\kappa$. This means that monitoring the system in photon counting simultaneously squeezes the projection noise during the time before a single photon is detected, in a manner very similar to the squeezing performed in phase-I. Therefore, the phase-II evolution of $\expval{\hat{X}_A^2}$ and $\expval{\hat{P}_A^2}$ up to the time of single photon detection, $t_2$, including optical pumping, is
\begin{align}
    \frac{\mathrm d}{\mathrm dt}\expval{\hat{X}_A^2}&=\frac{\kappa N}{16}e^{-4\gamma (t+t_1)}-2\gamma\expval{\hat{X}_A^2}+\gamma ,\label{eq:x2_w_dec_SPD}\\
    \frac{\mathrm d}{\mathrm dt}\expval{\hat{P}_A^2}&=-\frac{\kappa N}{4}\expval{\hat{P}_A^2}^2-2\gamma\expval{\hat{P}_A^2}+\gamma.\label{eq:p2_w_dec_SPD}
\end{align}
The effect of this evolution is to reduce the fluctuations in the $\hat P_A$ quadrature, which decreases the probability of seeing a single photon during some small time interval the longer one waits. 

During phase-I and phase-II, just before the detection of a single photon, the state is Gaussian, which we denote $\hat\rho_\text{pre}$.  Setting the mean to zero, the state is specfied by the variances after times $t_1, t_2$, $\expval{\hat{X}_A^2}(t_1,t_2)$ and $\expval{\hat{P}_A^2}(t_1,t_2)$, which can for find exactly by solving Eqs.~(\ref{eq:x2_w_dec}), (\ref{eq:p2_w_dec}), and (\ref{eq:x2_w_dec_SPD}), (\ref{eq:p2_w_dec_SPD}). These solutions can be found in App. \ref{app:x2_p2}. Using these, the Wigner function immediately before photon detection at $t_2$ is 
\begin{equation}
    W_\text{pre}(X,P,t_1,t_2)=\frac{\exp[-\frac12\left(\frac{X^2}{\expval{\hat{X}_A^2}(t_1,t_2)}+\frac{P^2}{\expval{\hat{P}_A^2}(t_1,t_2)}\right)]}{2\pi\sqrt{\expv{\hat{X}_A^2}(t_1,t_2)\expval{\hat{P}_A^2}(t_1,t_2)}}.\label{eq:W_pre}
\end{equation}
This is generally a squeezed state.

Given this Gaussian state $\hat\rho_\text{pre}$ prior to a photon click, the posterior state conditioned on detection of a single photon at time $t_2$ is given by application of the Kraus operator Eq. (\ref{eq:KSPD_dt}) according to
\begin{equation}
    \hat\rho_\text{post}=\frac1{\expval{\hat P_A^2}}\hat P_A\hat\rho_\text{pre}\hat P_A.
\end{equation}
In the presence of optical pumping, we can write the Kraus operator in an interaction picture \cite{Forbes_2024} as a differential operator (known as the \textit{Bopp representation}~\cite{Polkovnikov2010}) on the Wigner function.  The Bopp representation for the operator acting of the Wigner function is
\begin{equation}
    (\hat P)_{\mathcal B}=P-e^{-2\gamma (t_1+t_2)}\frac{i}{2}\frac{\partial}{\partial X}\equiv\mathcal P.\label{eq:P_correspondence}
\end{equation}
where the subscript $\mathcal B$, and similarly its complex conjugate when action of the right. $\mathcal{P}$ is a time dependent operator that includes the effect of decoherence that occurs due to optical pumping as discussed in ~\cite{Forbes_2024}. The post measurement state Wigner function is thus
\begin{equation}
    W_\text{post}(X,P,t_1,t_2)=\frac{1}{\expval{P^2}(t_1,t_2)}\mathcal P\mathcal{P^*}W_{\text{pre}}(X,P,t_1,t_2),\label{eq:W_final}
\end{equation}
The post-measurement state is a nonGaussian mixed state.

\begin{figure}[t]
    \centering
    \includegraphics[width=\linewidth]{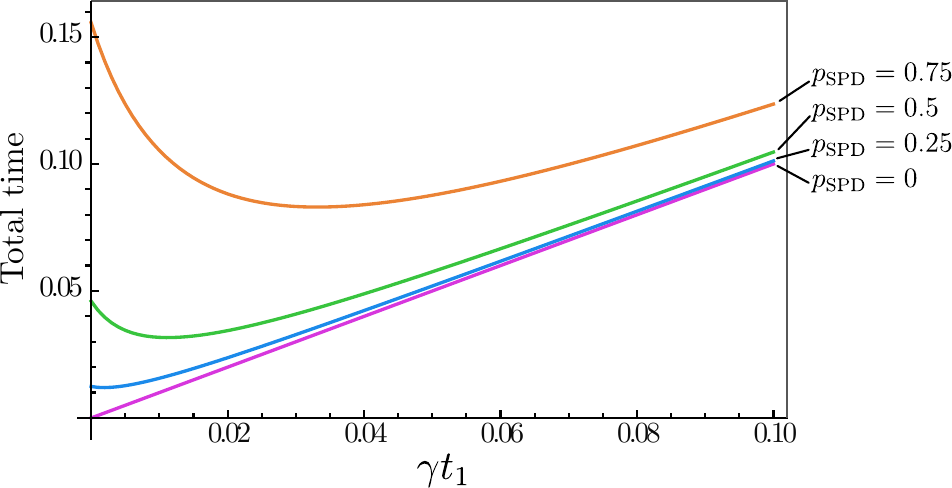}
    \caption{Plot of the total time $T=t_1+t_2$ of the experiment for several post-selection probability thresholds $p_\text{SPD}$. Parameters are $\kappa=\gamma$ and $N=500$. The time spent in phase-II, $t_2$, is made a function of both $t_1$ (the $x$-axis) and $p_\text{SPD}$. We highlight that for larger $p_\text{SPD}$ a minima in $T$ occurs as a function of $t_1$, indicating that there exists an optimal time to squeeze the ensemble if one wishes to minimze the total time $T$.}
    \label{fig:total_time}
\end{figure}

Finally, we calculate the total measurement time, $T=t_1+t_2$, as a function of $t_1$, the amount of time spent in phase-I. As discussed above, the probability of observing a single photon in phase-II, $p_{\text{SPD}}$, is a function of the of the variance in $\hat P_A$ [Eq.~(\ref{eq:p_spd})], and thus $t_2$ is a function of $t_1$ for a given post-selection probability. In order to minimize the total time $T$, one should set $t_1$ based on the probability with which they hope to observe a photon on each run. As seen in Fig.~\ref{fig:total_time}, if one post-selects only on very rare events of single photon detection during phase-II ($p_{\text{SPD}}=0$), the total time is monotonic with $t_1$. However, for higher probability thresholds, $p_\text{SPD}$, a minimum occurs in the plot of $T(t_1)$.  Minimizing the total runtime of the experiment limits the amount of decoherence that occurs from start to finish, and therefore may be the most applicable figure of merit in situations where decoherence plays a dominating role.

\subsection{Finite Detection Efficiency}
\label{sec:detection_efficiency}
The finite detection efficiency of photon counting implies that a photon that is collectively scattered into the forward direction is not counted.  This leads to collective decoherence, separate from optical pumping due to local scattering by one of the atoms into $4\pi$ steradians. In this section we will update our equations of motion to include the effects of this noise channel.

During homodyne and single photon detection, some portion of signal photons emitted by the ensemble will not be detected. This could be due to several reasons, including finite efficiency from the detector itself, absorption from the beam splitters and optics, or mode mismatch. All of these lead to collective decoherence and affect the state of the atoms similarly. We will define the fraction of signal photons which make it to the detector to be $\eta$. Therefore this decoherence can be modeled by sending measurement rate $\kappa\to\eta\kappa$ in our current equations of motion, and by adding the evolution caused by not detecting the signal. 

Beginning with homodyne detection, we apply the unitary map Eq.~(\ref{eq:bosonic_unitary}) to the joint atom-light initial state and trace out the light to obtain
\begin{equation}
    \Tr_L(\hat U\hat \rho_A\otimes\hat\rho_L\hat U^\dagger)=\hat \rho_A+\mathcal L\left[\sqrt{\frac{\kappa N_A\delta t}{8}}\hat P_A\right]\hat \rho_A,\label{eq:traced_evol}
\end{equation}
where $\mathcal L[\;\cdot\;]$ is a Lindbladian map with jump operator equal to its input. This evolution occurs at a rate $(1-\eta)\kappa$, and therefore from Eq.~(\ref{eq:traced_evol}) we see that the dissipative evolution of the atom subsystem is
\begin{equation}
    \frac{\mathrm d}{\mathrm dt}\hat\rho_A=\frac{\kappa(1-\eta) N_A}{8}\left[\hat P_A\hat \rho_A\hat P_A-\frac12\hat P_A^2\hat\rho_A-\frac12\hat\rho_A\hat P_A^2\right].\label{eq:hom_open_evol}
\end{equation}
This evolution on a Gaussian state $\hat \rho_A$ leads to equations of motion (App. \ref{app:x2_p2})
\begin{align}
    \frac{\mathrm d}{\mathrm dt}\expval{\hat{X}_A^2}&=\frac{\kappa(1-\eta)N}{8}e^{-4\gamma t}\expv{\hat{X}_A^2},\\
    \frac{\mathrm d}{\mathrm dt}\expval{\hat{P}_A^2}&=0.
\end{align}
Thus, the total evolution during homodyne detection is
\begin{align}
    \frac{\mathrm d}{\mathrm dt}\expval{\hat{X}_A^2}&=\frac{\kappa N}{8}e^{-4\gamma t}-2\gamma\expval{\hat{X}_A^2}+\gamma\label{eq:x2_w_dec_w_eta},\\
    \frac{\mathrm d}{\mathrm dt}\expval{\hat{P}_A^2}&=-\frac{\kappa \eta N}{2}\expval{\hat{P}_A^2}^2-2\gamma\expval{\hat{P}_A^2}+\gamma.\label{eq:p2_w_dec_w_eta}
\end{align}
Note that the detection efficiency only affects the rate of squeezing, but not the rate at which the variance in $\hat X_A$ increases.  

Including the same Lindblad evolution with the equations of motion from single photon detection we obtain
\begin{align}
    \frac{\mathrm d}{\mathrm dt}\expval{\hat{X}_A^2}&=\frac{\kappa(1-\eta/2) N}{8}e^{-4\gamma (t+t_1)}-2\gamma\expval{\hat{P}_A^2}+\gamma,\label{eq:full_phase_II_a}\\
\frac{\mathrm d}{\mathrm dt}\expval{\hat{P}_A^2}&=-\frac{\kappa \eta N}{4}\expval{\hat{P}_A^2}^2-2\gamma\expval{\hat{X}_A^2}+\gamma.\label{eq:full_phase_II_b}
\end{align}
We note that in this case the dependence on $\eta$ in the evolution of $\expval{\hat{X}_A^2}$ does not cancel.

\section{Figures of Merit}
\label{sec:figures_of_merit}

To analyze the utility of these states for quantum information processing and their susceptibility to decoherence, we study three possible figures of merit. We begin by studying the classical Fisher information (CFI) with respect to sensing a rotation about the $\hat J_z$ axis and measuring its displacement along $\hat J_y\propto \hat X_A$. We then study the quantum Fisher information (QFI) which quantifies the maximum CFI one can achieve by the best possible POVM used to sense the rotation given the state we prepare.  We deduce a measurement basis which comes close to achieving the maximum CFI and thus saturating the quantum Cram\'{e}rRao bound~\cite{braunstein1994statistical}.

\subsection{Classical Fisher Information}
In single parameter estimation theory one considers a probability distribution $p(x;\theta)$, where $x$ is a random variable and $\theta$ parameterizes some transformation on the distribution. The goal is to estimate the value of $\theta$ by repeatedly sampling from the distribution $n$ times. The Cram\'er-Rao bound sets a limit on the ultimate achievable variance of the estimator of $\theta$,
\begin{equation}
    \Delta\theta_{\text{est.}}^2\geq\frac1{n\mathcal F_C[p(x;\theta),\theta]},
\end{equation}
where $\mathcal F_C[p(x;\theta),\theta]$ is the Fisher information. In this context we will refer to this as the classical Fisher information (CFI) in order to distinguish it from the quantum Fisher information (QFI) discussed below.

The single parameter CFI of a probability distribution $p(x;\theta)$ is defined as the variance of the score of $p(x;\theta)$,
\begin{equation}
    \mathcal F_C[p(x;\theta),\theta]=\text{Var}\left[\frac{\partial}{\partial\theta}\ln(p(x;\theta))\right],\label{eq:cfi_def}
\end{equation}
which applies to both discrete and continuous probability distributions $p(x;\theta)$. Given an efficient, unbiased estimator one is guaranteed that the Cram\'er-Rao bound can be asymptotically saturated~\cite{pezze2018quantum}. Thus, the CFI is typically considered a relevant figure of merit when the goal is to sense transformations which can be parameterized by $\theta$.

\begin{figure}
    \centering
    \includegraphics[width=\linewidth]{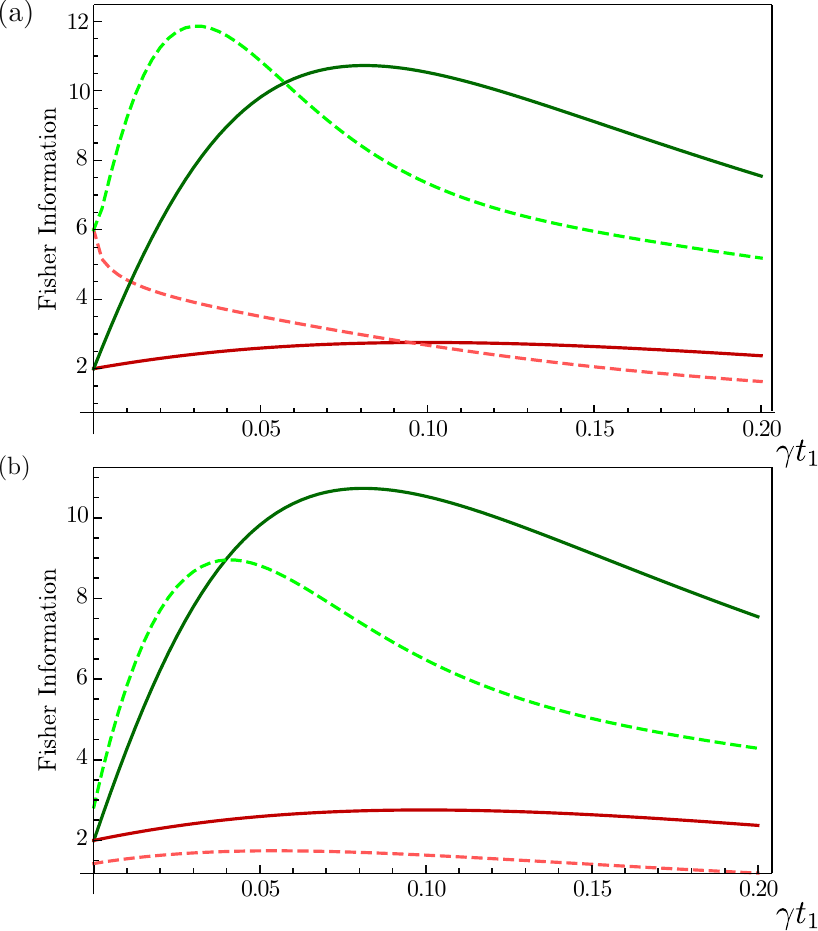}
    \caption{ Classical Fisher information (CFI) associated with sensing rotations by measuring spin projection along the direction of rotation. We plot the CFI as a function for time $t_1$ associated with phase-I of the protocol given the state created by measurement backaction and decoherence by optical pumping.  The solid lines are for phase-I only and the dashed lines include phase-II. Plot parameters for both (a) and (b) are $\kappa=\gamma$, $N=500$, and $\eta=1$. The green lines show CFI for $\kappa=\gamma$, while the red lines are plotted for $\kappa=0.1 \gamma$. (a) We post-select on cases where a single photon is detected immediately after phase 2 begins, i.e., $t_2=0$. This corresponds to the ``best case scenario,'' as less time has elapsed during which anti-squeezing and decoherence will affect the state. (b) The value of $t_2$ is adjusted for each value of $t_1$ such that the probability of seeing a single photon reaches at least $20\%$.}
    \label{fig:cfi}
\end{figure}

When sensing rotations of large ensembles of atoms, a natural geometry is to use spins that are polarized along the direction perpendicular to axis of rotation and to measure the spin projection in the direction of the rotation. The probability distribution in this example is
\begin{equation}
    p(m_y;\theta)=\bra{m_y}e^{-i\theta\hat J_z}\hat\rho e^{i\theta\hat J_z}\ket{m_y},
\end{equation}
where $\hat\rho$ is the probe state polarized along the $J_x$ direction, and $\ket{m_y}$ are eigenstates of $\hat J_y$. In the HPA this distribution becomes
\begin{equation}
    p(X;\theta)=\bra X\hat D_x(\theta)\hat\rho_\text{HP}\hat D^\dagger_x(\theta)\ket X,
\end{equation}
where $\hat D_X(\theta)$ is a displacement in the $X_A$ direction by an amount $\theta$, and $\hat\rho_\text{HP}$ is equivalent bosonic state in the HPA.

The simplest way to gain metrological advantage is to squeeze the projection noise along the direction of displacement. As the squeezed state is a Gaussian state, for a displacement along $X_A$, measurement in the $\ket X$ basis is optimal, meaning $p(X;\theta)$ achieves the maximum possible CFI of all measurement bases. For nonGaussian states one may achieve Fisher information above that of a squeezed state. While the ideal protocol we presented will achieve this, the decoherence that accompanies the measurement process can act to reduce the advantage and ultimately degrade the state so that hybrid measurement protocol is counterproductive.

To analyze this we calculate the CFI as a function of time $t_1$ in which the homodyne measurement is performed. During phase-I the measurement backaction of the detected signal induces squeezing, but optical pumping leads to depolarization of the mean spin and excess quadrature noise.  In the HPA, and for the proper choice of frame, this noise is represented as excess noise in both quadratures, as well as time dependent damping in the operators (see Eq. (\ref{eq:P_correspondence}) in ref. \cite{Forbes_2024}). In this bosonic representation, the noise due to optical pumping and depolarization of the mean spin appears as excess fluctuations in the anti-squeezed quadrature; the bosonic state is mixed and no longer minimum uncertainty.  In this case, photon subtraction/addition will not necessarily increase the CFI for the displacement measurement. We thus compare the CFI using the nonGaussian state, $\hat\rho_\text{NG}$, obtained by our protocol to that of a Gaussian squeezed state, $\hat\rho_\text{G}$, attainted solely by phase-I homodyne detection.

The results of this comparison are shown in Fig.~\ref{fig:cfi}. Solid lines are the CFI associated with homodyne measurement of the displaced Gaussian state for phase-I only, $\hat\rho_\text{G}$, and dashed lines are the CFI associated with the displaced nonGaussian state including phase-II (photon subtraction/addition), $\hat\rho_{\text{NG}}$ .  The green and red lines represent different measurement rates relative to the rate of optical pumping, $\kappa=\gamma$ and $\kappa=0.1\gamma$ respectively. We consider two scenarios of post-selection on single photon detection.  In Fig.~\ref{fig:cfi}(a), we post-select on the case that a single photon is detected immediately after phase-I (i.e., $t_2=0$).  This is the best-case scenario, because as $t_2$ increases the no-jump evolution reduces the CFI.   For comparison we consider post-selecting by choosing $t_2$ such that the probability of detecting a photon reaches at least $20\%$ (this threshold is arbitrarily chosen as a proof-of-principle to demonstrate the effect of waiting for a detection). In Fig.~\ref{fig:cfi}(b) the CFI is plotted with the same parameters as Fig.~\ref{fig:cfi}(a), but with $t_2$ adjusted at each value of $t_1$ to obtain at least $20\%$ detection probability.  One observes in Fig.~\ref{fig:cfi}(b) that even for large measurement strength $\kappa\sim\gamma$ (green), the need to wait for a detection drastically reduces the Fisher information gained from photon detection. For a lower measurement strength, $\kappa=0.1\gamma$ (red), there is no range of value of $t_1$ for which detection provides an increase in Fisher information.

The classical Fisher information with respect to sensing rotations informs us that while an improvement in precision can be gained by applying single photon detection, this improvement is severely limited by practical limitations of the experiment, like finite measurement rate $\kappa$ compared to the rate of optical pumping and finite detection time $t_2$. When one further takes detection efficiency $\eta<1$ into account, the improvement disappears entirely.  However, the CFI is not the true measure of quantum advantage.  As we will show below, the proper measurement to detect the displacement of the nonGaussian mixed state can yield substantial quantum advantage over squeezing alone. 

\subsection{Quantum Fisher Information}
Given a probe state used for sensing, the measurement (POVM) one should perform to best deduce an unknown parameter is not a priori determined. The quantum Fisher information (QFI) quantifies the largest possible CFI one can obtain by performing an optimal POVM, and using the measurement outcomes in an unbiased estimator.  When an unknown parameter $\theta$ is encoded in a state via a unitary transformation generated by a Hermitian operator $\hat A$, $\hat{\rho}_\theta = e^{-i\theta \hat{A}}\hat{\rho}e^{+i\theta \hat{A}}$,  the QFI can be written as
\begin{equation}
    \mathcal{F}_Q=2\sum_{\lambda,\lambda'}\frac{(\lambda-\lambda')^2}{\lambda+\lambda'}|\mel{\lambda}{\hat A}{\lambda'}|^2,
\end{equation}
where $\{\lambda,\ket\lambda\}$ are the eigenvalues and eigenvectors respectively of $\hat{\rho}_\theta$. 

In this section we will again consider the family of states generated by displacements in the bosonic representation
\begin{equation}
    \hat\rho(\theta)=\hat D_x(\theta)\hat\rho_\text{NG}\hat D^\dagger_x(\theta),
\end{equation}
where $\hat\rho_\text{NG}$ is the nonGaussian state generated upon single photon detection. We will compare the CFI of measuring this state in the position basis (homodyne detection) to the QFI, and present a near optimal POVM that comes close to saturating the quantum Cram\'{e}r-Rao bound.

\begin{figure}
    \centering
    \includegraphics[width=\linewidth]{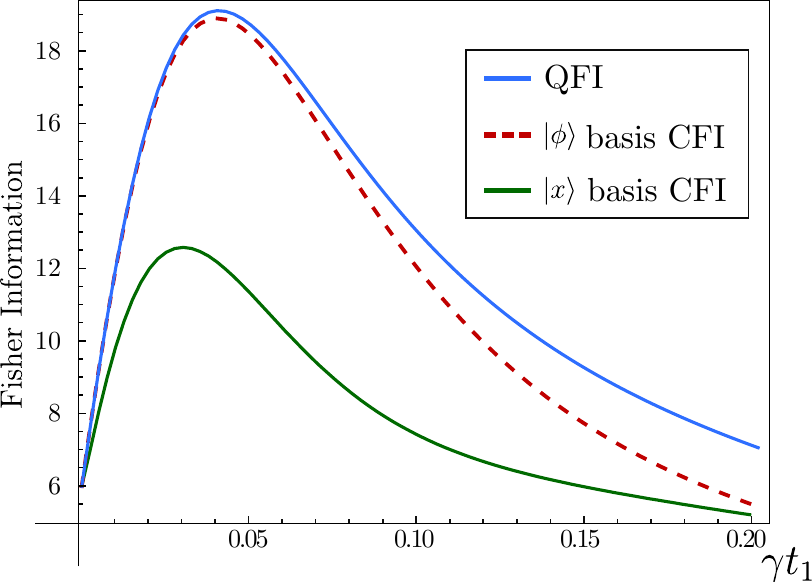}
    \caption{Fisher information for photon subtracted states as a function of the time spent applying homodyne measurement ($t_1$), with $N=500$, $\kappa=\gamma$, $t_2=0$, and $\eta=1$. The green line is the quantum Fisher information, the blue line is the classical Fisher information associated with a position basis measurement, and the red-dashed line is the classical Fisher information associated with measuring in the eigenbasis of $\hat P^{-1}\hat X\hat P^{-1}$.}
    \label{fig:qfi}
\end{figure}

In Fig.~\ref{fig:qfi} we plot the CFI for homodyne detection and the QFI as a function of $t_1$, the amount of time we squeeze through measurement-induced backaction in phase-I. The CFI (green) is significantly less than the QFI (blue), and thus a homodyne measurement of the post-measurement spin ensemble is far from optimal when detecting a rotation in the presence of noise.

Given the gap between the CFI and QFI, we seek a different POVM that makes use of the nonGaussianity of this mixed state for detecting displacements. Since the state of interest is proportional to $\hat{P}_A\hat\rho_\text{G}\hat{P}_A$, and $\hat\rho_\text{G}$ is optimally measured in the eigenbasis of $\hat{X}_A$, we explore a measurement in the eigenbasis of $\hat\Phi \equiv \hat{P}_A^{-1}\hat{X}_A\hat{P}_A^{-1}$. This is analogous to the measurement Lodschimdt echo procedure to measure the optimal basis for Ramsey spectroscopy with GHZ states~\cite{leibfried2004toward}. The eigenvalues of $\hat\Phi$ are continuous, and the corresponding eigenvectors are (App.~\ref{app:deriving_phi}),
\begin{equation}
    \ket\phi=\frac1{\sqrt{2\pi}}\int_{-\infty}^\infty pe^{-(i/3)p^3\phi}\ket p\;\mathrm dp,\label{eq:phi_definition}
\end{equation}
which are Dirac-delta normalized,
\begin{align}
    \braket{\phi'}{\phi}&=\frac1{2\pi}\int_{-\infty}^\infty p^2e^{-(i/3)p^3(\phi-\phi')}\mathrm dp\\
    &=\frac1{2\pi}\int_{-\infty}^\infty e^{-i u(\phi-\phi')}\mathrm du\\
    &=\delta(\phi-\phi').
\end{align}
These states also form a resolution of the identity
\begin{align}
    \int\ketbra\phi\mathrm d\phi=\mathds1,
\end{align}
making them a proper POVM.

\begin{figure*}[t]
    \centering
    \includegraphics[width=\linewidth]{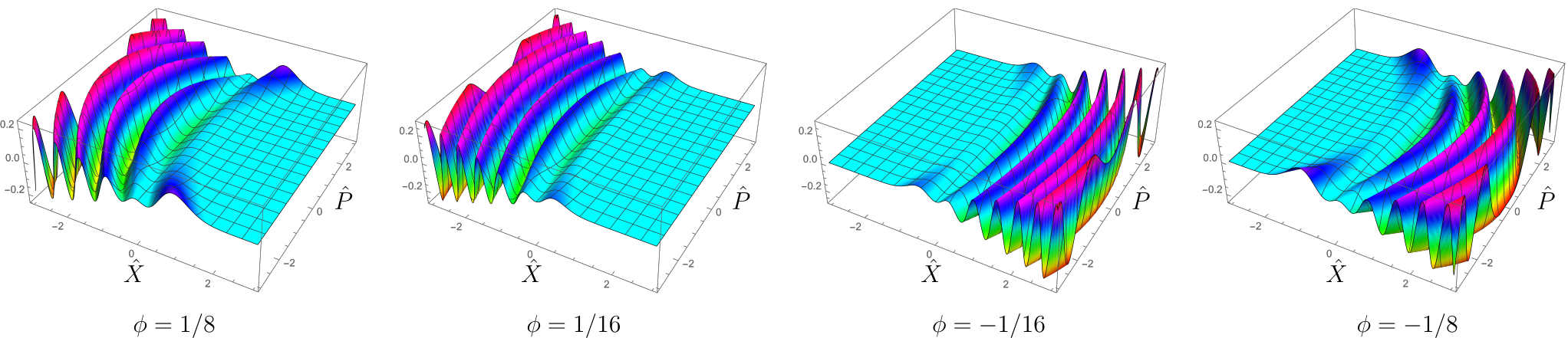}
    \caption{Wigner functions of the states $\ket\phi$, Eq. (\ref{eq:phi_definition}), for $\phi=1/8,1/16,-1/16,$ and $-1/8$ from left to right.}
    \label{fig:wigner_function}
\end{figure*}

The states $\ket\phi$ form a continuous-variable measurement basis, and thus share many of the same properties as $\ket x$ and $\ket p$. Illuminating examples include the fact that the overlap with a normalized state $\braket{\phi}{\psi}$ can be greater than 1, and that $\ket\phi$ is not a physically possible state, as no physical state can be dirac delta normalized. Despite these facts, $\mathrm d\phi \ketbra\phi$ forms a proper operator-valued measure on the space, and may be attainable in some limiting cases, similar to measurements in the $\ket X$ and $\ket P$ basis.  The Wigner function of the state $\ketbra\phi$ is plotted in Fig.~\ref{fig:wigner_function} and is explicitly given by (see App.~\ref{app:deriving_phi})
\begin{equation}
    W_\phi(X,P)=\frac{X}{\pi\phi(2\phi)^{1/3}}\text{Ai}\left[(2X-2\phi P^2)/(2\phi)^{1/3}\right],
\end{equation}
where Ai$(x)$ is the Airy function, arising from the cubic phase dependence in the definition of $|\phi\rangle$. Fringes in this measurement basis are essential for achieving the quantum advantage in sensing with the nonGaussian state.  

As shown in Fig.~\ref{fig:qfi}, in this measurement basis the CFI as a function of $t_1$ is very near the optimal measurement at early times, and only begins to deviate as the Fisher information decays.   In principle, thus, one can obtain substantial quantum advantage in the two-phase hybrid measurement.

\section{Summary and Outlook}
\label{sec:conclusions}
In this article we proposed a protocol to generate an approximate spin-squeezed Dicke state of an atomic spin ensemble using the measurement-backaction realized by entangling light with the atoms, and performing both Gaussian (homodyne) and nonGaussian (photon counting) measurements of the light. This hybrid-measurement approach both allows for the production of highly nonclassical states and an increase in the measurement strength for a given ensemble.  Using the Holstein-Primakoff approximation for large ensembles, and our previously derived master equation to include the inevitable decoherence that occurs by optical pumping~\cite{Forbes_2024}, we showed that the resulting mixed nonGaussian state can yield substantial quantum advantage for detecting spin rotations beyond that using spin-squeezed coherent states.  We quantified this advantage via the QFI. We showed that in the presence of optical pumping, the QFI associated with sensing a rotation of the final state is not achieved by the usual homodyne measurement, and that the discrepancy between the CFI of a homodyne measurement and the QFI is quite large. This shows the importance of searching for optimal measurement strategies in the presence of noise, which can yield substantial improvement over the strategy one would take in the absence of noise.  This was recently demonstrated in the context of Ramsey spectroscopy of GHZ-states in the presence of decoherence~\cite{kielinski2024ghz}.

We found a POVM that is close to an optimal measurement for sensing rotations with the mixed state prepared in our hybrid measurement scheme. In particular we considered a basis consisting of the eigenstate of the Hermitian operator $\hat{P}_A^{-1}\hat{X}_A \hat{P}_A^{-1}$s, denoted $\ket{\phi}$. We chose this in the spirit the Lodschmidt echo protocol used for implementing optimal measurements in Ramsey spectroscopy with GHZ states~\cite{leibfried2004toward}. An important difference between this measurement and the Lodschmidt echo protocol is that we measure in the $\hat{P}_A^{-1}\hat{X}_A \hat{P}_A^{-1}$ basis as a method to make our measurement more robust to noise. By comparison, in the protocol used in \cite{leibfried2004toward} the ``echo'' is used to implement a more complex measurement using more practical ones. The POVM we found from the eigenstates of $\hat{P}_A^{-1}\hat{X}_A \hat{P}_A^{-1}$ is a continuous-variable unnormalizable basis, similar to the position or momentum eigenstates. Their phase-space representation are Airy functions, which themselves are related to ``cubic phase states'' studied in continuous variable quantum computing and the original GKP proposal~\cite{gottesman2001encoding}, as seen in Eq.~(\ref{eq:phi_definition}). 

Finally, we showed that the total runtime of the experiment becomes a nonmonotonic function of $t_1$, the time in phase-I when we perform Gaussian squeezing, and when post-selecting on detecting a single photon with a chosen probability. This means that there exists optimal times to run phase-I if one wishes to decrease the total time of the experiment.  At longer times the deleterious effects of decoherence dominate over the benefit of measurement backaction.

An key outcome of this work is to demonstrate the fragility of an optimal POVM for metrology in the presence of decoherence.  In the ideal case, for a Gaussian state, or squeezed Fock state, the optimal measurement for detecting a rotation (displacement) is to measure the projection along the direction of displacement.  However, this measurement is far from optimal for when the state is subjected to certain noise channels.  In the future we plan to study how the CFI changes when the prepared state is imperfect.  This has important implications for the robustness of measurement strategies for sensing in the presence of decoherence.

\section{Acknowledgments}
The authors thank F. Elohim Becerra for his helpful discussions on this protocol, and Marco A. Rodr\'{i}guez-Garc\'{i}a for insights into the quantum Fisher information and sensing. This work is supported by funding from the NSF Quantum Leap Challenge Institutes program, Award No. 2016244.

\appendix

\section{Solving for $\expval{\hat X_A^2}$ and $\expval{\hat P_A^2}$ under Gaussian measurements and decoherence.}
\label{app:x2_p2}

In Sec.~\ref{sec:homodyne} we found the evolution of variance of the atomic quadratures due to measurement backaction arising from homodyne detection of the light. In Sec.~\ref{sec:SPD} we found differential equations that describe the evolution of the variance of the quadratures during the time in which we monitor the state, but have not yet detected a signal photon. Finally, in Sec.~\ref{sec:detection_efficiency} we included finite detection efficiency $\eta$ in our calculations. In this Appendix we will explicitly derive the equations of motion of $\expval{\hat X_A^2}$ and $\expval{\hat P_A^2}$ using the Kraus operators, and then provide full solutions to the differential equations (\ref{eq:x2_w_dec_w_eta})-(\ref{eq:full_phase_II_b}) for an initial spin coherent state.

We begin by applying the Kraus operator for homodyne measurement in Eq.~(\ref{eq:K_phase1}) to an arbitrary initial state $\hat\rho$ for small $T=\mathrm dt$ to obtain
\begin{align}
    \hat\rho+\mathrm d\hat\rho&=\hat K\hat\rho\hat K^\dagger+C\label{eq:KrhoK}\\
    &=\hat\rho-\mathrm dt\frac18\kappa N_A\{\hat P_A^2,\hat\rho\}+\mathrm dt\frac14\kappa N_A\expval{\hat P_A^2},
\end{align}
where the constant $C$ arises due to renormalization of the state. It follows that the evolution of $\hat\rho$ is governed by,
\begin{equation}
    \frac{\mathrm d}{\mathrm dt}\hat\rho=-\frac18\kappa N_A\{\hat P_A^2,\hat\rho\}+\frac14\kappa N_A\expval{\hat P_A^2}.\label{eq:drho_homodyne}
\end{equation}
We note that Eq.~{\ref{eq:drho_homodyne}} is equivalent to the ``no-jump'' evolution of $\hat\rho$ with respect to nonHermitian effective Hamiltonian associated with the jump operator $\hat L=(1/4)\kappa N_A\hat P_A$. This fact will become important later. 

To derive equations of motion for $\expval{\hat X_A^2}$ and $\expval{\hat P_A^2}$ we first turn Eq.~(\ref{eq:drho_homodyne}) into a differential equation on the Wigner function $W(X,P)$ by making using of Bopp representation \cite{Polkovnikov2010}. We do so by applying the formalism in \cite{Forbes_2024} to write $\hat P_A$ as
\begin{equation}
    (\hat P_A)_{\mathcal B}=P-e^{-2\gamma t}\frac i2\frac{\partial }{\partial X},\label{eq:P_bopp}
\end{equation}
where the subscript $\mathcal B$ indicates that this is the Bopp representation of the operator $\hat P_A$. The appearance of $e^{-2\gamma t}$ in Eq.~(\ref{eq:P_bopp}) is the result of going an appropriate frame~\cite{Forbes_2024}. This should be thought of as a sort of interaction picture, useful for describing the spin ensemble in the Holstein-Primakoff approximation in the presence of optical pumping. This picture is useful in the sense that it preserves constant proportionality between moments of the distribution $W$ in the bosonic mode, and moments of $\hat J_y$ and $\hat J_z$ in the spin system.

Applying Eq.~(\ref{eq:P_bopp}) to Eq.~(\ref{eq:drho_homodyne}) we obtain
\begin{align}
\frac{\mathrm d}{\mathrm dt}W&=-\frac{\kappa N}{8}\left[2P^2-\frac12e^{-4\gamma t}\frac{\partial^2}{\partial X^2}\right]W+\frac{\kappa N}{4}\expval{P^2}W\\
&=\left[-\frac{\kappa N}{4}P^2+\frac{\kappa N}{4}\expval{P^2}\right]W+\frac{\kappa N}{16}e^{-4\gamma t}\frac{\partial^2}{\partial X^2}W,\label{eq:W_evol_homodyne}
\end{align}
and integrating both sides of Eq.~(\ref{eq:W_evol_homodyne}) with $X^2$ and $P^2$ we obtain
\begin{align}
    \frac{\mathrm d}{\mathrm dt}\expval{P^2}&=-\frac{\kappa N_A}{4}\expval{P^4}+\frac{\kappa N_A}{4}\expval{P^2}^2\label{eq:p2_unfinished}\\
    \frac{\mathrm d}{\mathrm dt}\expval{X^2}&=\frac{\kappa N}{8}e^{-4\gamma t}.
\end{align}
This yields the desired answer for $\expval{X^2}$, but we can further simplify Eq.~(\ref{eq:p2_unfinished}) by enforcing that $W$ is Gaussian. Doing so we obtain
\begin{equation}
    \frac{\mathrm d}{\mathrm dt}\expval{P^2}=-\frac{\kappa N_A}{2}\expval{P^2}^2.
\end{equation}
Finally, one may add to these equations the evolution caused by optical pumping \cite{Forbes_2024} to obtain Eqs.(\ref{eq:x2_w_dec}) and (\ref{eq:p2_w_dec}). 

To derive the equations of motion for $\expval{X^2}$ and $\expval{P^2}$ during single photon detection prior to detection we note that the Kraus operator in Eq.~(\ref{eq:KSPD}) is  very similar to that used in homodyne detection Eq.~(\ref{eq:K_phase1}), but with half the rate $\kappa$. Therefore the derivation for the evolution of $\expval{X^2}$ and $\expval{P^2}$ during this phase-Is identical to the derivation above, but with $\kappa\to\kappa/2$ as seen in Eqs. (\ref{eq:x2_w_dec_SPD}) and (\ref{eq:p2_w_dec_SPD}).

Finally, we include finite detection efficiency $\eta$ by adding Eq.~(\ref{eq:hom_open_evol}) to Eq.~(\ref{eq:drho_homodyne}) to obtain
\begin{align}
    \frac{\mathrm d}{\mathrm dt}\hat\rho=&\eta\left[-\frac18\kappa N_A\{\hat P_A^2,\hat\rho\}+\frac14\kappa N_A\expval{\hat P_A^2}\right]\nonumber\\
    &+(1-\eta)\frac{\kappa N_A}{8}\left[\hat P_A\hat \rho_A\hat P_A-\frac12\hat P_A^2\hat\rho_A-\frac12\hat\rho_A\hat P_A^2\right].
\end{align}

Converting this into an equation of motion on a Wigner function $W$ using the correspondence rule in Eq.~(\ref{eq:P_bopp}) we obtain
\begin{align}
    \frac{\mathrm d}{\mathrm dt}W=&\left[-\frac{\kappa \eta N}{4}P^2+\frac{\kappa\eta N}{4}\expval{P^2}\right]W+\frac{\kappa\eta N}{16}e^{-4\gamma t}\frac{\partial^2}{\partial X^2}W\nonumber\\
    &+\frac{\kappa(1-\eta) N}{16}e^{-4\gamma t}\frac{\partial^2}{\partial X^2}W\\
    =&\left[-\frac{\kappa \eta N}{4}P^2+\frac{\kappa\eta N}{4}\expval{P^2}\right]W+\frac{\kappa N}{16}e^{-4\gamma t}\frac{\partial^2}{\partial X^2}W.\label{eq:W_with_eta}
\end{align}
We find that the second derivative term in Eq.~(\ref{eq:W_with_eta}) loses it dependence on $\eta$, while the other terms become proportional to $\eta$. Physically this means that the diffusion of the Wigner function in the $\hat X$ direction occurs at a constant rate regardless of $\eta$, but the amount of squeezing in the $\hat P$ direction can be decreased by $\eta$. For $\eta=0$ Eq.~(\ref{eq:W_with_eta}) only contains diffusion, as expected. In order to obtain Eqs.(\ref{eq:x2_w_dec_w_eta}) and (\ref{eq:p2_w_dec_w_eta}) one simply needs to integrate Eq.~(\ref{eq:W_with_eta}) with $X^2$ and $P^2$ respectively.

In order to derive Eqs. (\ref{eq:full_phase_II_a}) and (\ref{eq:full_phase_II_b}) for evolution during single photon detection we follow a similar procedure. We combine the evolution in Eq.~(\ref{eq:hom_open_evol}) with the 
 evolution in Eq.~(\ref{eq:drho_homodyne}), but with the rate $\kappa$ sent to $\kappa/2$ in the latter. Doing so we find that the equation of motion for $\hat\rho$ during this phase of the experiment is
\begin{align}
    \frac{\mathrm d}{\mathrm dt}\hat\rho=&\eta\left[-\frac1{16}\kappa N_A\{\hat P_A^2,\hat\rho\}+\frac18\kappa N_A\expval{\hat P_A^2}\right]\nonumber\\
    &+(1-\eta)\frac{\kappa N_A}{8}\left[\hat P_A\hat \rho_A\hat P_A-\frac12\hat P_A^2\hat\rho_A-\frac12\hat\rho_A\hat P_A^2\right].
\end{align}
The evolution of $W$ is therefore
\begin{align}
    \frac{\mathrm d}{\mathrm dt}W=&\left[-\frac{\kappa \eta N}{8}P^2+\frac{\kappa\eta N}{8}\expval{P^2}\right]W+\frac{\kappa\eta N}{32}e^{-4\gamma t}\frac{\partial^2}{\partial X^2}W\nonumber\\
    &+\frac{\kappa(1-\eta) N}{16}e^{-4\gamma (t_1+t)}\frac{\partial^2}{\partial X^2}W\\
    =&\left[-\frac{\kappa \eta N}{8}P^2+\frac{\kappa\eta N}{8}\expval{P^2}\right]W\nonumber\\
    &+\frac{\kappa (1-\eta/2) N}{16}e^{-4\gamma (t_1+t)}\frac{\partial^2}{\partial X^2}W.\label{eq:W_phaseII_all}
\end{align}

The addition of the $t_1+t$ term comes from the fact that time dependence in Eq.~(\ref{eq:P_bopp}) is in reference to the total time of the experiment, which during phase-II is equal to $t_1+t$. In Eq.~(\ref{eq:W_phaseII_all}) we see that $\eta$ now not only affects the squeezing, but also the diffusion. The lower the detection efficiency $\eta$, the faster the Wigner function diffuses in the $\hat X$ direction. This is in contrast to Eq.~(\ref{eq:W_with_eta}), in which the diffusion is independent of $\eta$.

The solutions for the variances,  $\expval{X^2}$ and $\expval{P^2}$, after the full protocol are
\begin{widetext}
\begin{align}
    \expval{X^2}=&\frac{1}{32\gamma}\bigg[e^{-4 \gamma  (t_2+ t_1)} \left((\eta -2) \kappa  N_A-(\eta -2) \kappa  N_A e^{2 \gamma  t_2}+\frac{16 \sqrt{2} \gamma  \left(2 \gamma ^2-\zeta ^2\right) e^{2 \gamma  (t_2+2  t_1)}}{\sqrt{2} \left(2 \gamma ^2+\zeta ^2\right)+4 \gamma  \zeta  \coth \left(\frac{\zeta   t_1}{\sqrt{2}}\right)}+16 \gamma  e^{4 \gamma  (t_2+ t_1)}\right)\bigg]\label{eq:x2_full}\\
    \expval{P^2}=&\bigg(32 \gamma  e^{4 \gamma   t_1} \bigg(-\frac{1}{8} e^{-4 \gamma   t_1} \left(\kappa   N-\kappa   N e^{2 \gamma   t_1}+8 \gamma  e^{4 \gamma   t_1}\right) \left(e^{\sqrt{\gamma } t_2\sqrt{4 \gamma +\eta  \kappa   N}}-1\right)\nonumber\\
    &-\frac{1}{16} \sqrt{\gamma } \sqrt{4 \gamma +\eta  \kappa   N} \left(\frac{\kappa   N e^{-4 \gamma   t_1} \left(e^{2 \gamma   t_1}-1\right)}{\gamma }+8\right) \left(e^{\sqrt{\gamma } t_2\sqrt{4 \gamma +\eta  \kappa   N}}+1\right)\bigg)\bigg)\nonumber\\
    &\bigg/\bigg(-\eta  \kappa ^2  N^2+\eta  \kappa ^2  N^2 e^{\sqrt{\gamma } t_2\sqrt{4 \gamma +\eta  \kappa   N}}-\eta  \kappa ^2  N^2 e^{\sqrt{\gamma } t_2\sqrt{4 \gamma +\eta  \kappa   N}+2 \gamma   t_1}+\eta  \kappa ^2  N^2 e^{2 \gamma   t_1}\nonumber\\
    &-8 \gamma  \left(8 \gamma +4 \sqrt{\gamma } \sqrt{4 \gamma +\eta  \kappa   N}+\eta  \kappa   N\right) e^{\sqrt{\gamma } t_2\sqrt{4 \gamma +\eta  \kappa   N}+4 \gamma   t_1}+8 \gamma  e^{4 \gamma   t_1} \left(8 \gamma -4 \sqrt{\gamma } \sqrt{4 \gamma +\eta  \kappa   N}+\eta  \kappa   N\right)\bigg).\label{eq:p2_full}
\end{align}
\end{widetext}

\section{Deriving $\ket\phi$}
\label{app:deriving_phi}

In this section we will demonstrate how to analytically solve for the eigenvectors of $\hat P^{-1}\hat X\hat P^{-1}$, where
\begin{equation}
    \hat P^{-1}=\int\frac1p\ketbra p\mathrm dp.
\end{equation}
We begin by noting that $\hat X$ can be written in the $\ket p$ basis as
\begin{align}
    \hat X&=\frac{1}{2\pi}\iiint x e^{-ixp}e^{ixp'}\ketbra{p}{p'}\mathrm dx\mathrm dp\mathrm dp'\\
    &=i\iint\delta'(p-p')\ketbra{p}{p'}\mathrm dp\mathrm dp',
\end{align}
where $\delta'(x)$ is the derivative of the Dirac delta distribution. Therefore
\begin{align}
    \hat P^{-1}\hat X\hat P^{-1}=i\iint\frac{1}{pp'}\delta'(p-p')\ketbra{p}{p'}\mathrm dp\mathrm dp'.
\end{align}

One can then find eigenvectors by defining an arbitrary state $\ket\phi$ in the $\ket p$ basis
\begin{equation}
    \ket\phi=\int c(p)\ket p\mathrm dp
\end{equation}
and calculating
\begin{align}
    \hat P^{-1}\hat X\hat P^{-1}\ket\phi&=i\iint\frac{c(p')}{pp'}\delta'(p-p')\ket p\mathrm dp\mathrm dp'\\
    &=i\int \frac{c(p)-pc'(p)}{p^3}\ket p\mathrm dp.
\end{align}
Setting this equal to $\phi\ket\phi$ leads to the differential equation
\begin{align}
    \phi c(p)=i\frac{c(p)-pc'(p)}{p^3},\label{eq:c_of_p_eq}
\end{align}
where $\phi$ is the eigenvalue of $\hat P^{-1}\hat X\hat P^{-1}$. Solving Eq.~(\ref{eq:c_of_p_eq}) we obtain
\begin{equation}
    \ket\phi=A\int pe^{-i\phi p^3/3}\ket p\mathrm dp,
\end{equation}
where $A$ is an arbitrary constant. Requiring that $\braket{\phi}{\phi'}=\delta(\phi-\phi')$ allows us to solve for $A=1/\sqrt{2\pi}$.

The Wigner function of these states can be found by first noting that
\begin{align}
    &(\ketbra\phi)_W\\
    &=\left(\hat P\iint e^{-i\phi p^3/3+i\phi' p'^3/3}\ketbra{p}{p'}\mathrm dp\mathrm dp'\hat P\right)_W\label{eq:pre_bopp}\\
    &=\left(p^2+\frac14\frac{\partial^2}{\partial x^2}\right)\left(\iint e^{-i\phi p^3/3+i\phi' p'^3/3}\ketbra{p}{p'}\mathrm dp\mathrm dp'\right)_W,\label{eq:bopp_on_wig}
\end{align}
where in Eq.~(\ref{eq:bopp_on_wig}) we have made use of the Bopp representation of $\hat P$. Solving for the Wigner function in Eq.~(\ref{eq:bopp_on_wig}) is much more tractable than the Wigner function in Eq.~(\ref{eq:pre_bopp}).

To solve for the Wigner function in Eq.~(\ref{eq:bopp_on_wig}) we calculate
\begin{widetext}
    \begin{align}
    W(x,\tilde p)&=\frac1\pi\int\mel{x-y}{\hat\rho}{x+y}e^{2ipy}\mathrm dy\\
    &=\frac{1}{2\pi^2}\int e^{-i\phi p^3/3+i\phi' p'^3/3}\braket{x-y}{p}\braket{p'}{x+y}e^{2ipy}\mathrm dy\mathrm dp\mathrm dp'\\
    &=\frac1{2\pi^2}\frac1{2\pi}\int e^{2i\tilde py}e^{-i\phi p^3/3+i\phi p'^3/3}e^{i(x-y)p-i(x+y)p'}\mathrm dp\mathrm dp'\mathrm dy\\
    &=\frac1{2\pi^2}\int e^{-i\phi p^3/3+i\phi p'^3/3}e^{ix(p-p')}\delta(2\tilde p-p-p')\mathrm dp\mathrm dp'\\
    &=\frac1{2\pi^2}\int e^{-i\phi p^3/3+i\phi (2\tilde p-p)^3/3}e^{ixp-ix(2\tilde p-p)}\mathrm dp\\
    &=\frac1{2\pi^2}\int e^{-i\phi (p+\tilde p)^3/3-i\phi(p-\tilde p)^3/3}e^{ix(p+\tilde p)}e^{ix(p-\tilde p)}\mathrm dp\\
    &=\frac1{2\pi^2}\int e^{-2i\phi (p^3+3\tilde p^2p)/3}e^{2ixp}\mathrm dp\\
    &=\frac1{2\pi^2}\int e^{-2i\phi p^3/3}e^{ip(2x-2\phi\tilde p^2)}\mathrm dp\\
    &=\frac1{2\pi^2(2\phi)^{1/3}}\int e^{-ik^3/3}e^{ik(2x-2\phi\tilde p^2)/(2\phi)^{1/3}}\mathrm dk\\
    &=\frac{1}{\pi(2\phi)^{1/3}}\text{Ai}\left[(2x-2\phi\tilde p^2)/(2\phi)^{1/3}\right],\label{eq:wigner_pre_bopp}
\end{align}
\end{widetext}
where in the last line we have used the fact that the Fourier transform of $e^{-ik^3/3}$ is the Airy function. Finally, applying the Bopp operator in Eq.~(\ref{eq:bopp_on_wig}) to Eq.~(\ref{eq:wigner_pre_bopp}) we obtain
\begin{align}
    W_{\ket\phi}(x,p)=&\left(p^2+\frac14\frac{\mathrm d^2}{\mathrm dx^2}\right)\nonumber\\
    &\times\frac{1}{\pi(2\phi)^{1/3}}\text{Ai}\left[(2x-2\phi p^2)/(2\phi)^{1/3}\right]\\
    =&\frac{x}{2\pi(2\phi)^{1/3}}\text{Ai}\left[(2x-2\phi p^2)/(2\phi)^{1/3}\right]
\end{align}

\bibliography{refs.bib}

\end{document}